\documentclass[5p]{elsarticle}

\usepackage{lineno}
\modulolinenumbers[5]

\usepackage{amsfonts}
\usepackage{algorithm, algorithmic}
\usepackage{epstopdf}
\usepackage{subcaption}
\usepackage{pgfplots}
\pgfplotsset{compat=1.7}
\usepackage{amsmath,stmaryrd,pict2e,picture}
\usepackage{float}

\begin{document}

\begin{frontmatter}

\title{Sparse Array Design for Maximizing the Signal-to-Interference-plus-Noise-Ratio by Matrix Completion}
\tnotetext[mytitlenote]{This work is supported by National Science Foundation (NSF) award no. AST-1547420.}

\author{Syed~A.~Hamza\corref{mycorrespondingauthor}}
\cortext[mycorrespondingauthor]{Corresponding author}
\ead{shamza@villanova.edu}
\author{Moeness~G.~Amin\corref{}}
\address{Center for Advanced Communications (CAC), Villanova University, PA 19085 USA}
\begin{abstract}
We consider sparse array beamfomer design achieving maximum signal-to interference plus noise ratio (MaxSINR).  Both array configuration and weights are attuned to the changing sensing environment.  This is accomplished by simultaneously switching among antenna positions and adjusting the corresponding    weights. The sparse array optimization design requires estimating the data autocorrelations at all spatial lags across the array aperture. Towards this end,  we adopt  low rank matrix completion under the semidefinite Toeplitz constraint for interpolating those autocorrelation values corresponding to the missing lags. We compare the performance of matrix completion approach with that of the fully augmentable sparse array design acting on the same objective function. The optimization tool employed is the regularized $l_1$-norm  successive convex approximation (SCA). Design examples with simulated data are presented using different operating scenarios, along with  performance comparisons among various configurations.
\end{abstract}

\begin{keyword}
Sparse arrays,  MaxSINR, SCA, Fully augmentable hybrid arrays, Matrix completion.
\end{keyword}

\end{frontmatter}

%\linenumbers

\section{Introduction}
Sensor selection schemes strive to optimize various performance metrics while curtailing   valuable hardware  and  computational resources. Sparse sensor placement, with various design objectives, has successfully been  employed   in  diverse application areas, particularly for enhanced parameter estimation and receiver performance \cite{710573, 6477161, 4058261, 6822510, 4663892, 5680595, 6031934, 8448767}.   The sparse array design criteria  are generally  categorized into environment-independent and  environment-dependent performance metrics. The  former are largely benign to the underlying environment and, in principle, seek to maximize the spatial degrees of freedom by extending the co-array aperture. This  enables high resolution direction of arrival (DOA) estimation possibly involving more sources than the available physical sensors \cite{1139138, 5456168, 7012090, 7106511, 8683315}. Environment-dependent objectives, on the other hand, consider the operating conditions characterized by emitters and targets in the array field of view, in addition to receiver noise. In this regard, applying such objectives  renders the array configuration as well as the array weights time-varying in response to dynamic and changing   environment.

In this paper, we focus on optimum sparse array design for receive beamforming that maximizes the output signal-to-interference and noise ratio (MaxSINR) \cite{1206680, 8378759, hamza2019sparse}.   It has been shown that optimum sparse array beamforming involves both array configuration and weights, and  can yield significant dividends in terms of SINR performance in presence of desired and interfering sources    \cite{6774934,  1634819, 6714077, 8061016, 8036237, 8313065, 7944303}.   However, one key challenge in implementing the data-dependent approaches, like Capon beamforming, is the need to have the exact or estimated values of the data autocorrelation function across the full sparse array aperture \cite{1206680, 1223538}. This underlying predicament arises as the sparse array design can only have few active sensors at a time, in essence  making it difficult to  furnish the correlation values corresponding to the inactive sensor locations. 

To address the aforementioned problem, we propose in this paper a  matrix completion strategy assuming a single desired source and multiple interfering sources.  This strategy permits the interpolation of the missing data correlation lags, thus enabling optimum “thinning” of the array for MaxSINR. The low rank matrix completion has  been utilized successfully in many applications, including the high-resolution direction of arrival estimation. We compare the matrix completion strategy to  the hybrid sparse array design that has been recently introduced and which also provides full spatial autocorrelation function for array thinning \cite{8682266}. The fundamental thrust of the hybrid design is to pre-allocate some of the available sensors  in such a way so as to ensure that all possible correlation lags can be estimated. In this case, the difference between the available sensors and those pre-allocated can  be utilized for maximum  SINR. In essence, the hybrid design locks  few spatial degrees of freedom in an attempt to making the full autocorrelation matrix  available to carry out the array optimization at all times. In that sense, it is a hybrid between structured and non-structured arrays. With pre-allocated sensors, the design approach  offers a simplified antenna switching as the environment changes.   In contrast, the matrix completion-based design   is not tied in to any pre-allocated sensor position and, therefore, has the ability to optimize over all the available sensor locations.  However, low rank matrix completion is a pre-processing step that is required every time we   decide on sensor selection as the environment changes.  This significantly  adds to the overall overhead and  computational complexity. We examine both approaches using estimated autocorrelation function, in lieu of its exact values, and compare their respective performances under different settings and degrees of freedom.

\begin{figure}[!t]
	\centering
	\includegraphics[width=3.47 in, height=2.25 in]{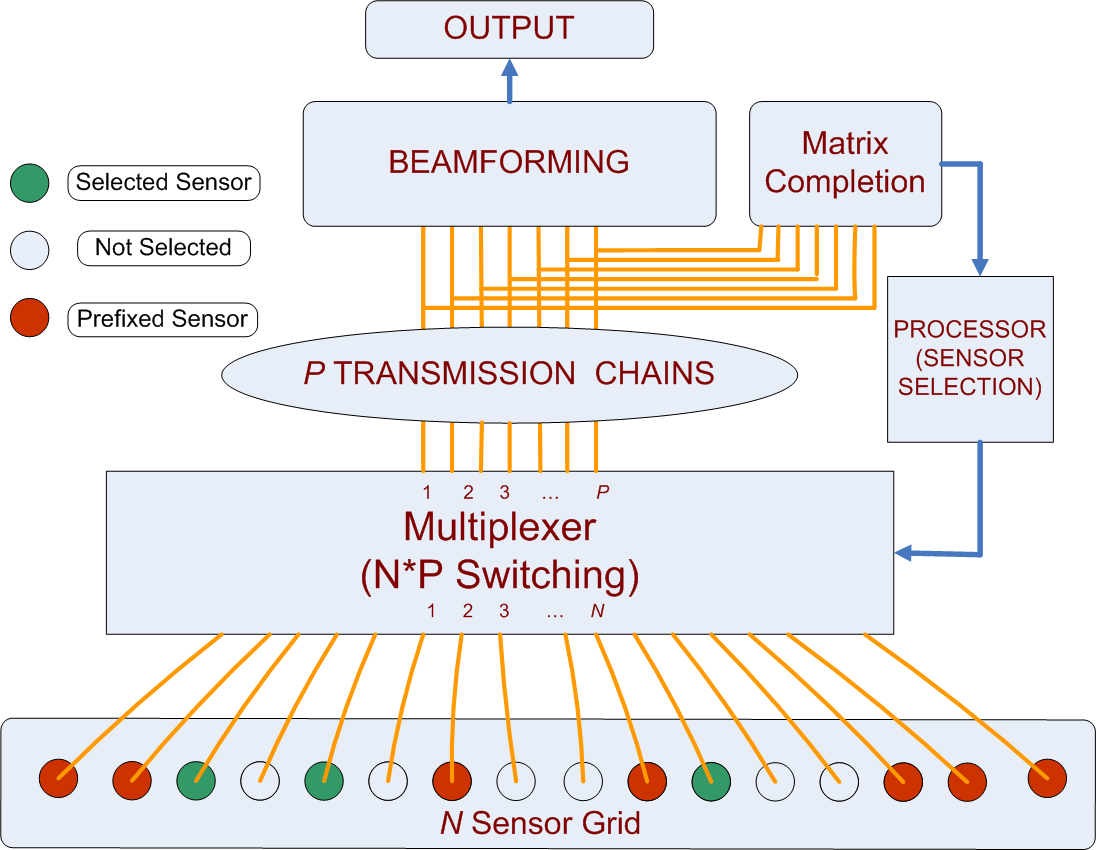}
	%where an .eps filename suffix will be assumed under latex,
	%and a .pdf suffix will be assumed for pdflatex; or what has been declare
	% via \DeclareGraphicsExtensions.
	\caption{Block diagram implementing adaptive beamforming and antenna switching}
	\label{Block_diagram}
\end{figure}

It is worth noting that  MaxSINR sparse array design using either  is an entwined optimization problem that  jointly optimizes the beamforming weights and determines the  active sensor locations. The   optimization  is posed as  finding   $P$ sensor positions out of $N$ possible equally spaced grid points for the highest  SINR performance. It is known that maximizing the SINR is equivalent to the problem of  maximizing  the  principal eigenvalue of the product of  the inverse of data correlation matrix and the desired source correlation matrix  \cite{1223538}. However, the maximum eigenvalue problem over all possible sparse topologies  is a combinatorial problem and is challenging to solve in polynomial times. To alleviate the computational complexity of exhaustive combinatorial search,  we pose this problem as successive convex approximation (SCA) with reweighted  $l_1$-norm regularization  to promote sparsity in the final solution.  

To proceed with  the SCA optimization, it is essential to input the algorithm with the full data correlation matrix. For the hybrid design, all the correlation lags are available  and, therefore,  we resort to averaging across the available correlation lags to harness a Toepltiz matrix estimate of the received data.  On the other hand, a sparse array designed freely without preallocating sensor locations necessitates the use of low rank matrix completion to interpolate the missing lags and subsequently applying the SCA optimization \cite{86455521}.  The word ``free" implies no preset position of any of the sensors involved. It is shown that the matrix completion  is  an effective approach to accomplish the MaxSINR sparse  design. The performance of matrix completion can potentially surpass the hybrid design at the expense  of more involved sensor switching and additional  computational complexity stemming from the Toeplitz interpolation of the missing correlation lags.

The rest of the paper is organized as follows:  In the next section, we  state the problem formulation for  maximizing the  output SINR. Section \ref{Optimum sparse array design} details  the SCA for arrays designed freely  alongside the     hybrid   design approach and the associated modified SCA optimization. Section \ref{sec4} explains the matrix completion approach.   In section \ref{Simulations}, with the aid of Monte Carlo simulations,  we compare the performance of  hybrid-designed arrays viz a viz freely-designed arrays in a limited snapshot environment. Concluding remarks follow at the end.

\section{Problem Formulation} \label{Problem Formulation}
We consider an emitter source in the presence of narrowband  interfering signals. The  signals impinge  on a  uniform grid of $N$ linear elements with the inter-element spacing  of $d$ and the received signal is  given by; 
\begin {equation} \label{a}
\mathbf{x}(n)=    b_s(n) \mathbf{s}( \theta)  + \sum_{k=1}^{I} b _{ik}(n) \mathbf{i}( \theta_k)  + \mathbf{v}(n),
\end {equation}
The sampling instance is  $n$, $I$ are the number of interfering sources and ($b_s(n)$, $b _{ik}(n))$ $\in \mathbb{C}$  are the baseband  signals  for the source and interferences, respectively. The  steering vector    corresponding to the  direction  of arrival of desired   source $\mathbf{s} ({\theta})$  $\in \mathbb{C}^N$ is    given by,
\vspace{+2mm}
\begin {equation}  \label{b}
\mathbf{s} ({\theta})=[1 \,  \,  \, e^{j (2 \pi / \lambda) d cos(\theta)  } \,  . \,   . \,  . \, e^{j (2 \pi / \lambda) d (N-1) cos(\theta)  }]^T.
\end {equation}
The interference steering vectors $\mathbf{i} ({\theta_k})$ are similarly defined. The additive  noise  $\mathbf{v}(n)$ $\in \mathbb{C}^N$ is Gaussian with variance    $\sigma_v^2$.
The beamformer processes the received signal  $\mathbf{x}(n)$  linearly   to improve   SINR. The beamformer output $y(n)$ is given by, 
\begin {equation}  \label{c}
y(n) = \mathbf{w}_o^H \mathbf{x}(n),
\end {equation}
The optimal beamforming weights $\mathbf{w}_o$ that maximizes SINR is given by solving the following optimization problem  \cite{1223538};
\begin{equation} \label{d}
\begin{aligned}
\underset{\mathbf{w} \in \mathbb{C}^N}{\text{minimize}} & \quad   \mathbf{w}^H\mathbf{R}_{i}\mathbf{w},\\
\text{s.t.} & \quad     \mathbf{ w}^H\mathbf{R}_{s}\mathbf{ w}=1.
\end{aligned}
\end{equation}
The desired source  correlation is  $\mathbf{R}_s=  \sigma^2 \mathbf{s}( \theta)\mathbf{s}^H( \theta)$, with source power $ \sigma^2 =E\{b_s(n)b_s^H(n)\}$. The sum of the  interference and uncorrelated noise correlation matrix is $\mathbf{R}_{i}= \sum_{k=1}^{I} (\sigma^2_k \mathbf{i}( \theta_k)\mathbf{i}^H( \theta_k)$) + $\sigma_v^2\mathbf{I}_{N\times N}$,    with the $k$th interference power $ \sigma^2_k =E\{b_{ik}(n)b_{ik}^H(n)\}$. Since  $\mathbf{R_{x}}=\mathbf{R}_s+ \mathbf{R}_{i}$, then formulation (\ref{d}) can be written as as part of the objective function as follows \cite{1223538},
\begin{equation} \label{e}
\begin{aligned}
\underset{\mathbf{w} \in \mathbb{C}^N }{\text{minimize}} & \quad   \mathbf{ w}^H\mathbf{R_{x}}\mathbf{ w},\\
\text{s.t.} & \quad     \mathbf{ w}^H\mathbf{R}_{s}\mathbf{ w} \geq 1.
\end{aligned}
\end{equation}
where the equality constraint is relaxed due to the inclusion of the relationship between the data and signal autocorrelation matrices in the cost function. The optimum solution of the above problem only  requires the knowledge of the received data correlation matrix $\mathbf{R_{x}}=E(\mathbf{x}\mathbf{x}^H)$ and the  DOA of the desired source. The former can readily be estimated from the received data vector $\mathbf{x}$ over $T$ snapshots, $\mathbf{\hat{R}_{x}}=\frac{1}{T}\sum_{n=1}^{T} \mathbf{x}(n)\mathbf{x}^H(n)$.

The analytical solution of the optimization problem    is given by $\mathbf{w}_o =  \{  \mathbf{R}_{i}^{-1} \mathbf{s}(\theta)  \}$ with the optimum output SINR$_o$;
\begin{equation}  \label{f}
\text{SINR}_o=\frac {\mathbf{w}_o^H \mathbf{R}_s \mathbf{w}_o} { \mathbf{w}_o^H \mathbf{R}_{i} \mathbf{w}_o} = \Lambda_{max}\{\mathbf{R}^{-1}_{i} \mathbf{R}_s\},
\end{equation}
which is in fact the maximum eigenvalue ($\Lambda_{max}$) of   the product of   inverse of data correlation matrix  and the desired source correlation matrix. In the next section, the  formulation in (\ref{e}) is extended to the sparse beamformer design.

\section{Sparse array design through SCA algorithm} \label{Optimum sparse array design}
The expression in (\ref{f}) is applicable to any array topology, including uniform and sparse arrays with the respective correlation matrices. To achieve sparse solutions, given the knowledge of full correlation matrix, (\ref{e}) is introduced with  an additional constraint,
\begin{equation} \label{a2}
\begin{aligned}
\underset{\mathbf{w \in \mathbb{C}}^{N}}{\text{minimize}} & \quad  \mathbf{ w}^H\mathbf{R_{x}}\mathbf{w},\\
\text{s.t.} & \quad   \mathbf{ w}^H\mathbf{R}_{s}\mathbf{ w}\geq1,  \\
& \quad   ||\mathbf{w}||_0=P.
\end{aligned}
\end{equation}
The  operator $||.||_0$ denotes the $l_0$ norm which constrains the cardinality of  the weight vector $\mathbf{w}$ to the number of available sensors, $P$.  The problem  in  (\ref{a2}) is clearly non convex   involving   a hard constraint, rendering the formulation challenging to solve in polynomial time \cite{doi:10.1002/cpa.20132}.

The objective function  and  quadratic constraint  in  (\ref{a2}) are interchanged,  transforming  into equivalent formulation as follows,
\begin{equation} \label{b2}
\begin{aligned}
\underset{\mathbf{w \in \mathbb{C}}^{N}}{\text{maximize}} & \quad   \mathbf{ w}^H\mathbf{R}_s\mathbf{ w},\\
\text{s.t.} & \quad     \mathbf{ w}^H\mathbf{R_x}\mathbf{ w} \leq 1. \\
& \quad   ||\mathbf{w}||_0=P.
\end{aligned}
\end{equation}
In general, the beamforming weight vector is complex valued, however the quadratic functions are real. The real and imaginary entries of the optimal weight vector are typically decoupled, permitting the involvements of only  real unknowns. This is achieved through concatenating the beamforming weight vector and defining  the respective correlation matrices   \cite{8445845},
\begin{multline} \label{c2}
\tilde{\mathbf{R}}_s=\begin{bmatrix}
\text{real}({\mathbf{R}}_s)       & -\text{imag}({\mathbf{R}}_s)   \\
\\
\text{imag}({\mathbf{R}}_s)       & \text{real}({\mathbf{R}}_s)   \\
\end{bmatrix},
\tilde{\mathbf{w}}=\begin{bmatrix}
\text{real}({\mathbf{w}})       \\
\\
\text{imag}({\mathbf{w}})   \\
\end{bmatrix}
\end{multline}
\begin{align} \label{d2}
\tilde{\mathbf{R}}_x=\begin{bmatrix}
\text{real}({\mathbf{R}}_x)       & -\text{imag}({\mathbf{R}}_x)   \\
\\
\text{imag}({\mathbf{R}}_x)       & \text{real}({\mathbf{R}}_x)   \\
\end{bmatrix}
\end{align}
Replacing $\mathbf{R_s}$ and $\mathbf{R_x}$ by $\tilde{\mathbf{R}_s}$ and $\tilde{\mathbf{R}_x}$ respectively, (\ref{b2}) can be expressed in terms of real variables,
\begin{equation} \label{e2}
\begin{aligned}
\underset{\mathbf{\tilde{w} \in \mathbb{R}}^{2N}}{\text{maximize}} & \quad   \mathbf{\tilde{ w}}^{'}\tilde{\mathbf{R}}_s\mathbf{\tilde{ w}},\\
\text{s.t.} & \quad     \mathbf{\tilde{ w}}^{'}\mathbf{\tilde{R}_x}\mathbf{\tilde{ w}} \leq 1. \\
& \quad   ||\mathbf{w}||_0=P.
\end{aligned}
\end{equation}
The quadratic constraint  clearly has the convex feasibility region, however, there is still a  non convex constraint involving the  $l_0$ norm.  In order to realize the convex feasible region,  the  $l_0$ norm is typically relaxed to the $l_1$ norm, which has been effectively used in many sparse recovery applications.  The maximization problem is first transformed to a minimization in order to move the $l_1$ norm constraint to the objective function and realize a sparse solution.    This is achieved   by reversing the sign of the entries of the desired source correlation matrix $\bar{\mathbf{R}}_s=-\mathbf{\tilde R}_s$,
\begin{equation} \label{f2}
\begin{aligned}
\underset{\mathbf{\tilde{w} \in \mathbb{R}}^{2N}}{\text{minimize}} & \quad   \mathbf{\tilde{ w}}^{'}\bar {\mathbf{R}}_s\mathbf{\tilde{ w}},\\
\text{s.t.} & \quad     \mathbf{\tilde{ w}}^{'}\mathbf{\tilde{R}_x}\mathbf{\tilde{ w}} \leq 1. \\
& \quad   ||\mathbf{w}||_0=P.
\end{aligned}
\end{equation}

To convexify the objective function, the concave objective  is  iteratively approximated through successive linear approximation,
\begin{equation} \label{g2}
\begin{aligned}
\underset{\mathbf{\tilde{ w} \in \mathbb{R}}^{2N}}{\text{minimize}} & \quad   \mathbf{m}^i{'}\mathbf{\tilde{ w}}+b^i,\\
\text{s.t.} & \quad     \mathbf{\tilde{ w}}^{'}\mathbf{\tilde{R}_x}\mathbf{\tilde{ w}}\leq 1. \\
& \quad   ||\mathbf{w}||_0=P.
\end{aligned}
\end{equation}
The approximation coefficients  $\mathbf{m}^i$ and ${b^i}$, are updated iteratively $\mathbf{m}^{i+1}=2\bar{\mathbf{R}}_s\mathbf{\tilde{ w}}^i, b^{i+1}=-\mathbf{\tilde{w}}^i{'}\bar{\mathbf{R}}_s \mathbf{\tilde{w}}^i $ by first order approximation. Finally, the non convex  $l_0$ norm is  relaxed  through minimizing the mixed $l_{1-\infty}$ norm to recover sparse solutions,
\begin{equation} \label{h2}
\begin{aligned}
\underset{\mathbf{\tilde{w} \in \mathbb{R}}^{2N}}{\text{minimize}} & \quad   \mathbf{m^i{'}}\mathbf{\tilde{ w}}+b^i + \mu(\sum_{k=1}^{N}||\mathbf{\tilde{w}}_k||_\infty),\\
\text{s.t.} & \quad     \mathbf{\tilde{ w}}^{'}\mathbf{\tilde{R}_x}\mathbf{\tilde{ w}}\leq 1.
\end{aligned}
\end{equation}
The summation implements the $l_1$ norm that is minimized as a convex  surrogate of $l_0$ norm. The vector $\mathbf{\tilde{w}}_k \in \mathbb{R}^{2}$ has two entries containing  the real and imaginary parts of the beamforming weight corresponding to the $k$th sensor. The $||.||_\infty$ selects  the maximum entry of  $\mathbf{\tilde{w}}_k$ and  discourages the real and imaginary entries concurrently. This is because not selecting a sensor  implies the simultaneous removal of both the real and corresponding imaginary entries in the final solution vector. The sparsity parameter  $\mu$ is set to zero for the first few initial iterations to allow the solution to converge to optimal solution for  the full array elements. The sparsity parameter $\mu$ by itself does not  guarantee   the final solution to be $P$ sparse. To guarantee a $P$ sparse solution, the optimization problem is solved successively against different values of $\mu$. The values of $\mu$  are typically given by a binary search  over the possible upper and lower limit of $\mu$ until the algorithm converges to $P$ sensors \cite{6477161}.  

\begin{algorithm}[t!] 
	
	\caption{SCA for  sparse array  beamforming.}
	
	\begin{algorithmic}[]
		
		\renewcommand{\algorithmicrequire}{\textbf{Input:}}
		
		\renewcommand{\algorithmicensure}{\textbf{Output:}}
		
		\REQUIRE Received data sparse correlation matrix $\mathbf{R}_P$, look direction DOA $\theta$. \\
		
		\ENSURE  $P$ sensor beamforming weight vector   \\
		
		\textbf{Matrix Completion:} \\
		 Run  Eq. (\ref{a3}) for free design and Toeplitz averaging for hybrid design to estimate the full correlation matrix.	\\	
		 Set the lowest eigenvalues of $\mathbf{\hat{R}_x}$ corresponding to the noise subspace to be equal to the noise floor.	\\	
		\textbf{Initialization:} \\
		Initialize the beamforming vectors randomly to find $\mathbf{m}$ and $b$. 
		Initialize  $\epsilon$, $\mu=0$.
		\WHILE {(Solution does not converge corresponding to $\mu=0$)}
		%		\WHILE {($|\mathbf{\tilde{W}|}$ $\leq (\text{threshold to  discard antenna })$)}
		
		\STATE   Run  Eq. (\ref{j2}).\\ 
		%		 Check if some entries in $\mathbf{\tilde{W}}$ is exactly zero, if yes, check\\ % Entering row contents\\
		\ENDWHILE \\
		Initialize $\mathbf{h}^{i}$= all ones vector, Binary vector for hybrid design.\\
		Select $\mu$ (Binary search)
		\WHILE {(Beamforming weight vector is not $P$ sparse)}
		%		\WHILE {($|\mathbf{\tilde{W}|}$ $\leq (\text{threshold to  discard antenna })$)}
		
		\STATE   Run  Eq. (\ref{j2})) . (for initial iteration use $\mathbf{m}^i$ and $b^i$ from  previous while loop)\\ 
		\STATE   Update the regularization weighting parameter,  $\mathbf{h}^{i+1}(k)=\frac{1}{||\mathbf{\tilde{w}}_k^i||_2+\epsilon}$, Update $\mathbf{m}^i$ and $b^i$ \\
		%		 Check if some entries in $\mathbf{\tilde{W}}$ is exactly zero, if yes, check\\ % Entering row contents\\
		\ENDWHILE \\
		
		\STATE After achieving the desired cardinality, analytically solve for  $\mathbf{\tilde{w}}$  corresponding to the selected sensor locations,  yielding, optimal weight vector. 
		
		\RETURN Optimal weight vector $\mathbf{w}_o$ 
	\end{algorithmic}
	\label{algorithm}
\end{algorithm}
\subsection{Hybrid sparse array design}
Formulation (14) penalizes all the sensor weights rather judiciously in an effort to  optimize the objective function. We refer to this approach as free-design. On the other hand, the hybrid sparse array design,  penalizes only some sensor weights, leaving the remaining sensors to assume prefixed positions. These position are chosen to guarantee full augmentatbility of the sparse array, i.e., provide the ability to estimate the autocorrelation at all spatial lags across the array aperture. This provide the means  for thinning the array and carrying out sparse optimization all the times.  In order to discriminate the  prefixed sensors from those which are available for design, the weighted formulation  is adopted, in turn modifying  (\ref{h2}) as follows, 
\begin{equation} \label{i2}
\begin{aligned}
\underset{\mathbf{\tilde{w} \in \mathbb{R}}^{2N}}{\text{minimize}} & \quad   \mathbf{m^i{'}}\mathbf{\tilde{ w}}+b^i + \mu(\sum_{k=1}^{N}\mathbf{h}(k)||\mathbf{\tilde{w}}_k||_\infty),\\
\text{s.t.} & \quad     \mathbf{\tilde{ w}}^{'}\mathbf{\tilde{R}_x}\mathbf{\tilde{ w}}\leq 1.
\end{aligned}
\end{equation}
The weighting vector $\mathbf{h}$ is a binary vector with $1's$ and $0's$ entries. The entries corresponding to the prefixed sensor locations are set to $0$ while the remaining entries are initialized to $1$. In this way, the partial penalization is implemented in (\ref{i2}) that ensures  the sparsity is not enforced to the prefixed locations. 
% \subsubsection{Enhancing Sparsity}
The weighted penalization can easily be extended to the reweighting  formulation which can further promote sparsity and facilitates the  $P$ sparse solution \cite{Candès2008}. This is achieved by iteratively updating weighting vector $\mathbf{h}$ \cite{6663667, article1},
\begin{equation} \label{j2}
\begin{aligned}
\underset{\mathbf{\tilde{w} \in \mathbb{R}}^{2N}}{\text{minimize}} & \quad   \mathbf{m^i{'}}\mathbf{\tilde{ w}}+b^i + \mu(\sum_{k=1}^{N}\mathbf{h}^i(k)||\mathbf{\tilde{w}}_k||_\infty),\\
\text{s.t.} & \quad     \mathbf{\tilde{ w}}^{'}\mathbf{\tilde{R}_x}\mathbf{\tilde{ w}}\leq 1.
\end{aligned}
\end{equation}
The re-weighting vector $\mathbf{h}^i$, at the $i$-th iteration, is updated as an inverse function of the beamforming weights at the present iteration,
\begin{equation} \label{k2}
\mathbf{h}^{i+1}(k)=\frac{1}{||\mathbf{\tilde{w}}_k^i||_2+\epsilon}
\end{equation}
This relatively suppresses the low magnitude weights in the next iteration to accelerate sparsity. The parameter $\epsilon$ avoids the case of division by zero. The reweighting is applied to both the freely designed  array and the hybrid design. For the former, the vector $\mathbf{h}$ is initialized to all $1's$ vector and updated iteratively. However, to preserve the prefixed sensor locations for the hybrid design, the entries of $\mathbf{h}^i$ corresponding to the prefixed locations must remain zero for all iterations, while the remaining entries are initialized to $1$   and updated as explained above. The procedure is summarized in Algorithm \ref{algorithm}.

\section{Toeplitz matrix completion and Fully augmentable completion through averaging} \label{sec4}
The key  concern in the free-design sparse array formulation is the assumption regarding the knowledge of the full array correlation matrix. This is because the data from only $P$ active  sensors is available to  estimate the  correlation matrix.  The full correlation matrix, in this case, is not  readily available and could have many missing correlation lags. Many different approaches for sparse matrix completion, under variant assumptions about the data model, have been considered in the literature including high resolution DOA estimation. We adopt a positive semidefinite Toepltiz  matrix completion scheme that effectively  exploits the structure of the unknown  correlation matrix. It is well known that the  narrowband far field  sources impinging on the ULA resultantly has  the  hermitian positive definite correlation matrix having the Toeplitz structure.    Along with the  Toepltiz  positive definite condition, the trace heuristic is incorporated to interpolate the missing lags. The trace heuristics is successfully used in many areas of control systems and array processing to recover simpler and low rank data models \cite{Recht:2010:GMS:1958515.1958520, Mohan:2012:IRA:2503308.2503351, 5447068}. Moreover, it has been shown  that the trace heuristic is equivalent to the nuclear norm minimization, rendering gridless recovery of the underlying narrowband sources, thus  recovering the missing correlation lags \cite{8474369, 8472789, 7539135, 7935534, 7833203}. The matrix completion problem is, therefore, written as,
\begin{equation} \label{a3}
\begin{aligned}
\underset{l \in \mathbb{C}^{N}}{\text{minimize}} & \quad  ||Toeplitz(l)\odot \mathbf{Z}-\mathbf{R}_P||_F^2 + \zeta \text{Tr}(Toeplitz(l))\\
\text{s.t.} & \quad   Toeplitz(l) \succeq 0  .
\end{aligned}
\end{equation}
Here,  $l$ is a complex vector with a real first element,  then $Toeplitz(l)$ returns the symmetric Toeplitz matrix having  $l$ and $l^H$ defining its first row and column  respectively. Matrix $\mathbf{R}_P$ is the received data correlation matrix with missing correlation lags. The entries corresponding to the missing correlation lags  are set to zero. The  symbol `$\odot$' denotes the element wise multiplication and  `$\succeq$' denotes the matrix inequality enforcing the positive  semidefinite constraint. The matrix  $\mathbf{Z}$  is a binary matrix which only fits the non zero elements in $\mathbf{R}_P$ to the unknown Toepltiz matrix.  The function `$||.||_F^2$' is the square of the Frobenius norm of the matrix which seeks to minimize  the sum of error square between the observed correlation values and the corresponding entries of the unknown Toepltiz matrix. The symbol `$\zeta$' gives the trade off  between the denoising  term and the trace heuristic pursuing simpler model. The nominal value of the  parameter  `$\zeta$'   is typically tuned from the numerical experience for the underlying problem.  However, the Toeplitz estimate can potentially be ill conditioned having quite a  few  eigenvalues close to zero.  We utilize the maximum likelihood estimate of the interpolated Toeplitz  correlation matrix by incorporating the knowledge of the noise floor. In  so doing, the eigenvalues corresponding to the noise subspace are set equal to the noise floor. 

Unlike the free-design sparse array, where missing lags manifest themselves as zero values at all entries of some of the autocorrelation matrix sub-diagonals, the hybrid design would ensure that at least one element in each matrix sub-diagonal is available. This facilities the Toeplitz estimation of the received data correlation matrix by averaging the non zero correlation entries across each sub-diagonal. The averaging scheme, however, does not guarantee the positive definiteness of the  Toeplitz estimate \cite{709534}, \cite{Abramovich:1999:PTC:2197953.2201686}. This renders the formulation in (\ref{j2}) non convex, which essentially requires $\mathbf{R_x}$ to be positive semidefinite. In order to circumvent this issue, we return to  the maximum likelihood estimate adopted for the matrix completion approach to facilitate a positive definite estimate by eliminating the negative eigenvalues typically appearing in the noise subspace. Finally, the estimated data correlation matrix $\mathbf{\hat R_x}=Toeplitz(l)$ is used in lieu of $\mathbf{R_x}$ to carry out the data dependent optimization for MaxSINR.

\section{Simulations} \label{Simulations}
We show   examples under different design scenarios to access  the  performance of the proposed methodology  achieving  MaxSINR.  We establish two  performance benchmarks in order to examine   the sensitivity of the proposed  algorithm  to  the initial array configuration. This is because the matrix interpolation approach is guided on the    initial configuration that decides the location of the  missing entries in the data correlation matrix. The initial  configuration refers to the $P$-element sparse array topology at the start before commencing of any adaptation process.   In general, the initial configuration could be any random array, or the optimized configuration from the preceding operating conditions.     The first performance benchmark applies the SCA algorithm under the assumption that the   data from  all the perspective  sensor locations  is available. In this way,  the actual  full  correlation matrix utilizing $T$  snapshots is input to the SCA algorithm. Clearly, the performance of the aforementioned  benchmark is not  reliant on the initial configuration  but is  dependent   on the observed data realization and  the number of   snapshots. Another  deterministic performance benchmark  assumes  perfect knowledge of the full correlation matrix, representing the case of  unlimited data snapshots. To draw a proper distinction, the former would be  referred as the ``Full correlation matrix-limited snapshots (FCM-LSS),'' and the latter is henceforth called the ``Full correlation matrix-unlimited snapshots (FCM-USS)''.  
%We demonstrate the  robustness of the design approach under the correlated jamming environment.
\subsection{Example comparing both designs}

\begin{figure}[!t]
	\centering
	\begin{subfigure}[b]{0.5\textwidth}
		\includegraphics[width=8.9cm, height=1.09cm]{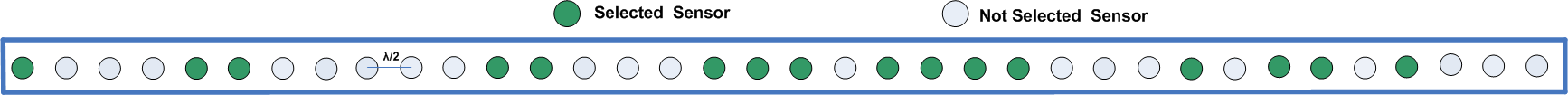}
		\caption{}
		\label{best_90_0}
	\end{subfigure}
	\begin{subfigure}[b]{0.5\textwidth}
		\includegraphics[width=8.9cm, height=1.09cm]{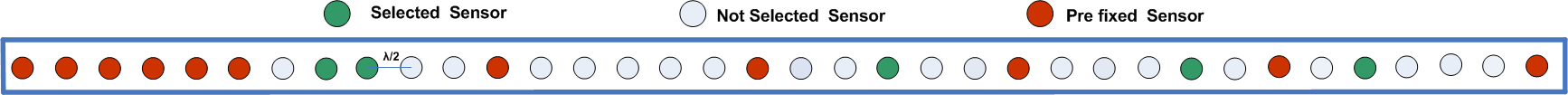}
		\caption{}
		\label{worst_90_0}
	\end{subfigure}
	\begin{subfigure}[b]{0.5\textwidth}
		\includegraphics[width=8.9cm, height=1.09cm]{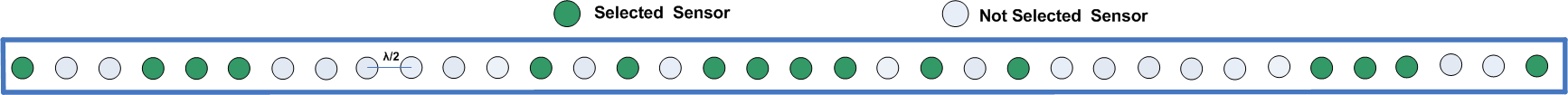}
		\caption{}
		\label{op_algo_90_0}
	\end{subfigure}
	\begin{subfigure}[b]{0.5\textwidth}
		\includegraphics[width=8.9cm, height=1.09cm]{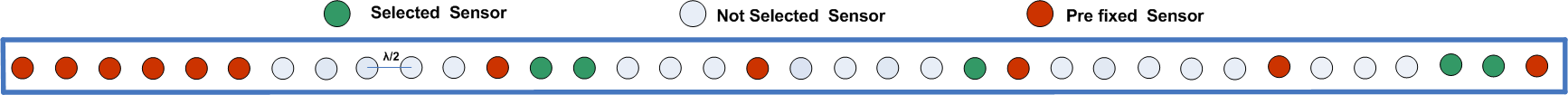}
		\caption{}
		\label{op_symalgo_90_0}
	\end{subfigure}
	\begin{subfigure}[b]{0.5\textwidth}
		\includegraphics[width=8.9cm, height=1.09cm]{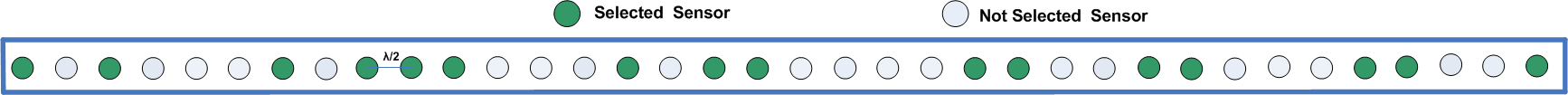}
		\caption{}
		\label{best_90_01}
	\end{subfigure}
	\begin{subfigure}[b]{0.5\textwidth}
		\includegraphics[width=8.9cm, height=1.09cm]{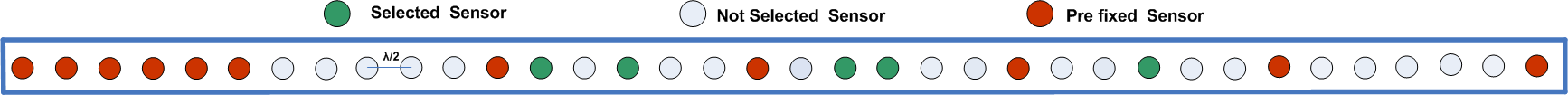}
		\caption{}
		\label{worst_90_01}
	\end{subfigure}
	\begin{subfigure}[b]{0.5\textwidth}
		\includegraphics[width=8.9cm, height=1.09cm]{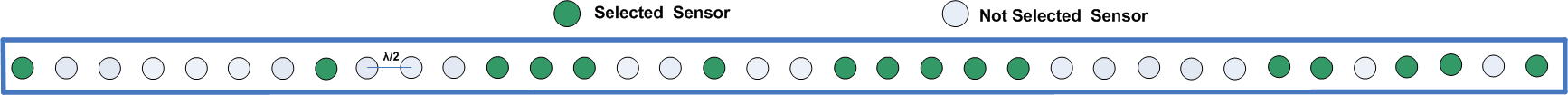}
		\caption{}
		\label{op_algo_90_01}
	\end{subfigure}
	\begin{subfigure}[b]{0.5\textwidth}
		\includegraphics[width=8.9cm, height=1.09cm]{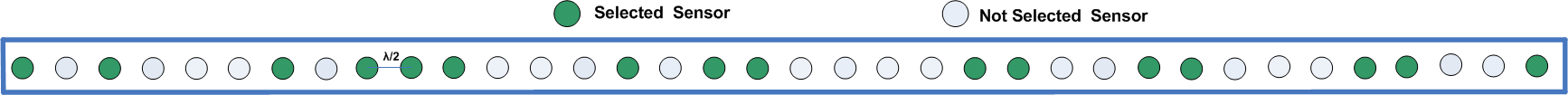}
		\caption{}
		\label{op_symalgo_90_01}
	\end{subfigure}
	\begin{subfigure}[b]{0.5\textwidth}
		\includegraphics[width=8.9cm, height=1.09cm]{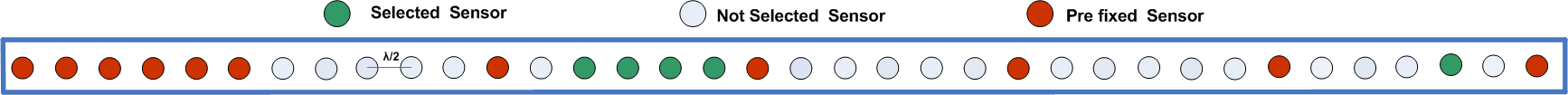}
		\caption{}
		\label{op_algo_90_02}
	\end{subfigure}
	\begin{subfigure}[b]{0.5\textwidth}
		\includegraphics[width=8.9cm, height=1.09cm]{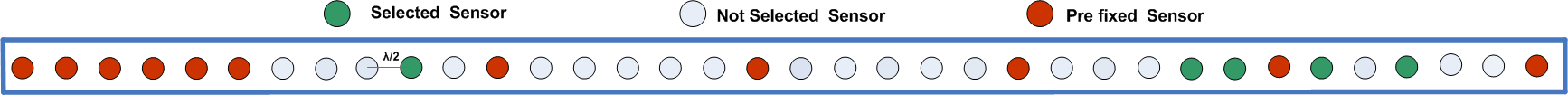}
		\caption{}
		\label{op_symalgo_90_02}
	\end{subfigure}
	\caption{(a) Initial   configuration; randomly selected 16 antennas from 36 (b) Initial   configuration leading to fully augmentable array   (c) Freely designed array (d) Hybrid designed array (e)  Initial random  configuration;  selected 16 antennas from 36 (f) Initial   configuration  leading to  fully augmentable array (g) Freely designed array (h) Hybrid designed array (i) Best performing array configuration  (j) Worst performing array configuration  }
	%\caption{(a) Optimum  (enumeration, source at $90^0$)  (b) Optimum (algorithm, (source at $90^0$))  (c) Worst array configuration (source at $90^0$) (d) Optimum (source at $40^0$) }
	\label{point}
\end{figure}
Given $N=36$ perspective   sensor locations placed linearly with an inter-element spacing of $\lambda/2$. Consider selecting $P=16$ sensors among these locations so as to maximize the SINR. A single  source of interest is  operating at   $90^0$, i.e. array broadside. There are also six jammers,   concurrently active at locations  $40^0$, $85^0$ $95^0$, $135^0$, $140^0$  and $160^0$. The SNR of the desired signal is $0$ dB, whereas each  jammer has the interference to noise ratio (INR)  of $20$ dB. The range of binary search for the sparsity parameter $\mu$ is set from $0.01$ to $5$, $\gamma=10^{-3}$ (sparsity threshold)  and $\epsilon=0.05$. The initial $16$-element sparse array configuration to estimate the data correlation matrix is randomly chosen, and shown in the    Fig. \ref{best_90_0}. This configuration has missing correlation lags and is occupying a fraction of the available aperture. The array collects the data for  $T=1000$ snapshots. The full array Toeplitz estimate is recovered through matrix completion with the regularization parameter $\zeta=0.5$.   The proposed SCA approach  employing  matrix completion renders an array configuration with SINR of $11.73$ dB. It is worth noting that, for the underlying case,  the number of possible array configurations is of  order   $10^9$ which makes the problem infeasible to solve  through exhaustive search. The upper bound of performance, however,  is  $12$ dB which corresponds to the case when interferences are completely canceled in the output. In this regard, the designed array configuration is very close to the performance upper bound. The optimized array configuration is shown in the  Fig. \ref{op_algo_90_0}. It is noted that this configuration has also missing few correlation lags.

In order to access the performance of the hybrid design approach, we  consider a randomly selected $16$ element fully augmentable array, which is shown in Fig. \ref{worst_90_0}. The full data correlation matrix is estimated using the same $T=1000$ snapshots and averaging is carried over the available correlation lags to yield a Toepltiz estimate. The SCA approach, in this case, achieves the array design shown in  Fig. \ref{op_symalgo_90_0} and has a reasonable SINR performance of $10.92$ dB. The designed hybrid array is fully augmentable and involves the  prefixed sensor locations which are arranged in the nested array topology (prefixed configuration shown in red color). The hybrid design is clearly sub optimal as compared to the array designed freely. It is noted that the number of possible  hybrid sparse array configurations associated with the prefixed sensors   is $53130$. Although, the possible fully augmentable configurations are significantly less as compared to $10^9$  possibilities, the maximum SINR hybrid design found through enumeration is $11.93$ dB and is close to the upper performance bound of $12$ dB.  The performance of both designs are compared with the benchmark design initialized with FCS-LSS  estimated from  $T=1000$ samples supposedly collected from all $N$ sensors. The benchmark design yields the freely designed and hybrid sparse configurations  with the  SINR of $11.82$ dB and $11.65$ dB respectively. This performance is  superior to the above mentioned designs that employ the Toeplitz estimation  in lieu of the actual  full correlation matrix. 

It is of interest to analyze the effect of the initial sparse array configuration on the proposed SCA optimization. This time, the data is collected through the initial configurations  depicted in Figs. \ref{best_90_01} and \ref{worst_90_01},  instead of the configurations (Figs. \ref{best_90_0} and \ref{worst_90_0}) employed for the earlier example. The underlying operating environment and all other parameters remain the same as above. As before, the freely designed array is achieved through matrix completion, whereas the hybrid design involves averaging to estimate the full data correlation matrix.  The free-design and the hybrid design achieve  SINR of $11.82$ dB and $11.65$ dB, respectively. The designed array configurations are shown in the  Figs. \ref{op_algo_90_01} and \ref{op_symalgo_90_01}.
These configurations offer superior performances to those optimized earlier, assuming different initial configurations. This  underscores the dependence of sparse array beamforming optimization on the array initial conditions. It is  noted that for the same underlying environment and  initial configuration, the proposed solution   is still not unique and dependent on the random realizations of the received data.  In order to reliably gauge the performance  of the proposed scheme, we report the average results   repeated over $100$ independent trials.  It is found that  under the initial configurations shown in Figs. \ref{best_90_0} and \ref{worst_90_0}, the average SINR performances are $11.79$ dB for freely designed SCA and $11.18$ dB for the hybrid design. On the other hand,   the initial configurations, shown in Figs. \ref{best_90_01} and \ref{worst_90_01}, yield  the average performances of $11.6$ dB   and $11.54$ dB for the free and hybrid designs, respectively. These performances are compared with the FCS-LSS benchmark.  It is found that  the  FCS-LSS  offers the same performance as is achieved by SCA under initial configurations adopted in Figs. \ref{best_90_01} and \ref{worst_90_01}.   We remark that under the initial array configurations shown in Figs. \ref{best_90_0} and \ref{worst_90_0}, the SCA-based matrix completion even surpasses the FCS-LSS benchmark, however, it offers slightly lower SINR for the hybrid design ($11.18$ dB as compared to $11.54$ dB). The optimum hybrid array configuration found through enumeration is shown in Fig. \ref{op_algo_90_02} with an SINR of $11.9$ dB, whereas the worst case hybrid configuration (shown in Fig. \ref{op_symalgo_90_02}) has an associated SINR of $7.5$ dB which is considerably lower than the above designs.

\subsection{Monte Carlo design for random scenarios}
\begin{figure}[!t]
	\centering
	\includegraphics[width=9.55cm, height=6.90cm]{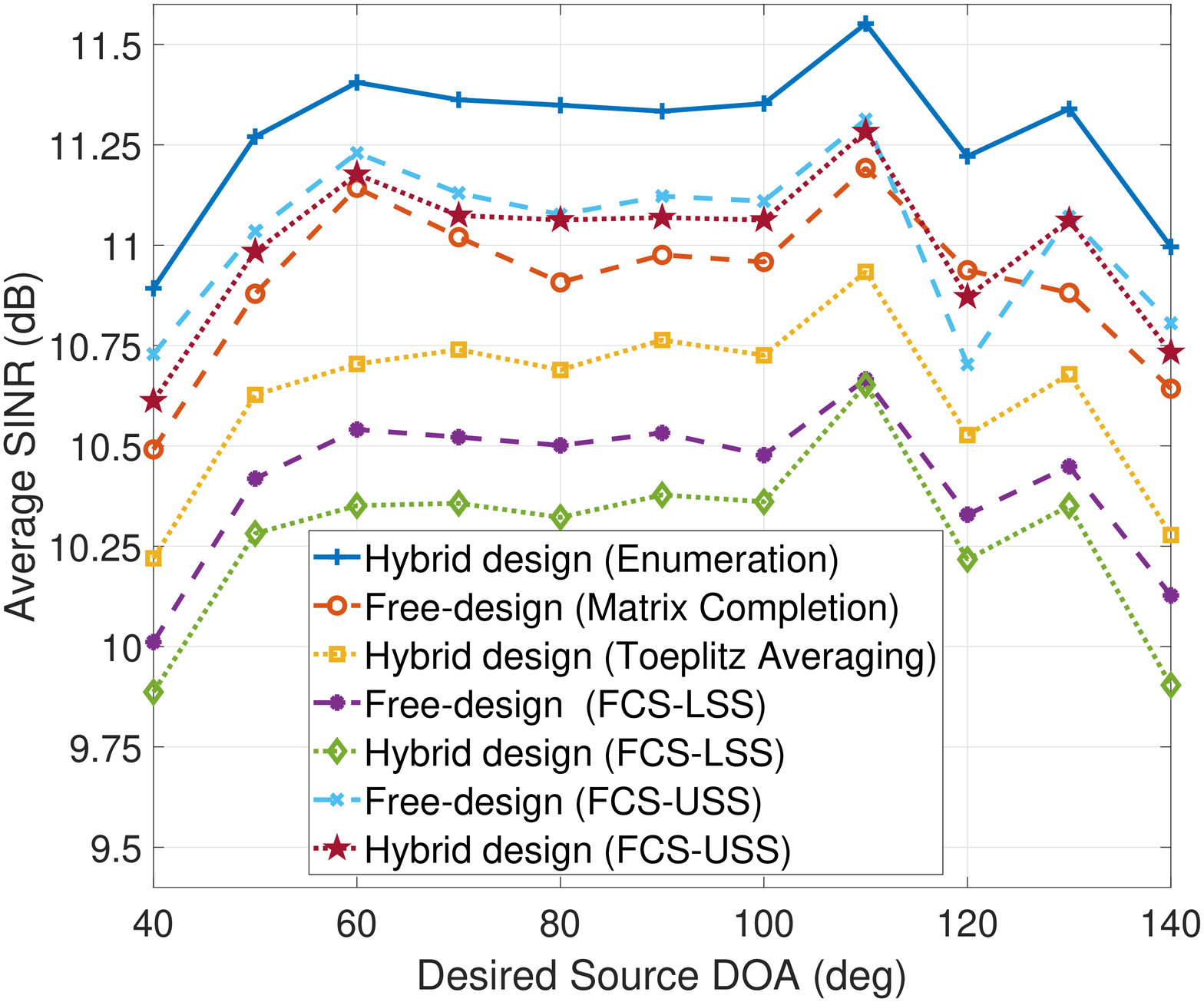}
	\caption{Average SINR performance of various sparse topologies against desired source DOA  for  $T=100$ snapshots.}
	\label{robust_bp1}
\end{figure}
\begin{figure}[!t]
	\centering
	\includegraphics[width=9.55cm, height=6.90cm]{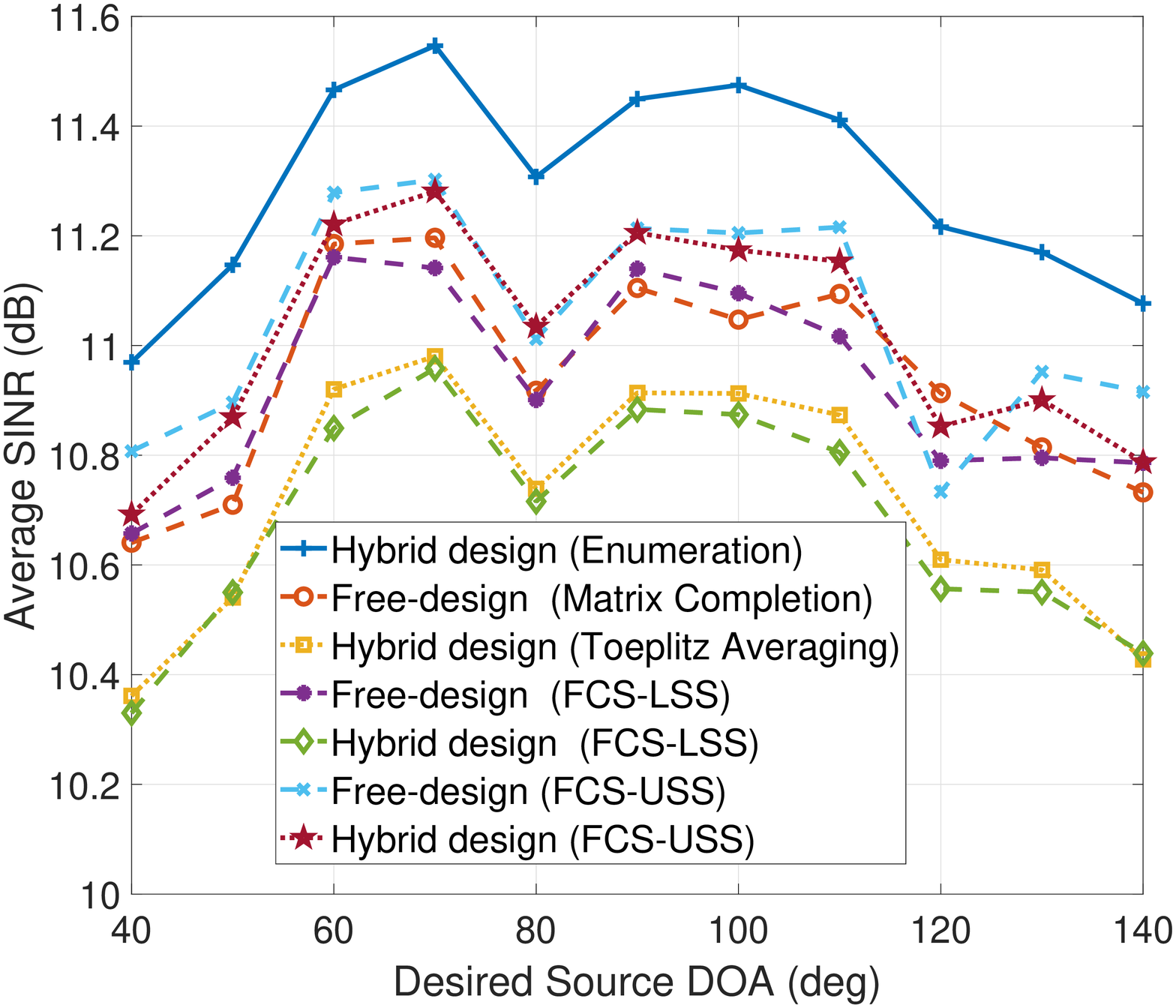}
	\caption{Average SINR performance of various sparse topologies against desired source DOA  for  $T=250$ snapshots.}
	\label{robust_bp2}
\end{figure}

The above examples tie the performance of the proposed algorithm  not only  to the location of the sources and their respective powers but  also show the dependence on the initial array configuration, the number of snapshots and the observed realization of the received data. In order to provide a more meaningful assessment, the simulation scenarios are designed  keeping the aforementioned variables in perspective. We generate $11$   different scenarios. For each scenario, the desired source DOA is kept fixed, whereas six jammers are randomly placed   anywhere from $30^0$ to $150^0$ with the respective powers uniformly distributed from $5$ to $15$ dB. The experiments are repeated $3000$ times and the initial array configuration is randomly chosen for each experiment. For the freely designed array, the initial array configuration is selected by randomly choosing $16$ sensors from $36$ sensors. However, the initial configuration for  the hybrid design is randomly chosen from all the possible $16$ sensor fully augmentable array configurations associated with the prefixed sensors arranged in nested configuration as depicted in  Fig. \ref{worst_90_0} (red color sensors). 

Figure \ref{robust_bp1}  shows the results for $T=100$. The performance curve of the SCA algorithm for the freely designed array incorporating  matrix completion lies in between (for most points) the benchmark designs incorporating   FCM-USS and  FCM-LSS.  That is the matrix completion approach  even outperforms  the benchmark design incorporating the FCM-LSS. This performance is explainable because  matrix completion coupled with the apriori knowledge of noise floor renders a more accurate estimate of the full correlation matrix as compared to  FCM-LSS,  without incorporating knowledge of noise floor, which has high noise variance because of limited snapshots. The performance of the other benchmark incorporating the exact knowledge of the  correlation matrix (FCM-USS) is clearly superior over  matrix completion. The results are fairly similar for the hybrid design, where the performance curve utilizing the Toeplitz averaging is sandwiched between the  benchmark designs incorporating the exact  correlation matrix (FCM-USS) and  the one  utilizing the  presumably observed full data correlation matrix (FCM-LSS). The hybrid designed and freely designed arrays, both demonstrate desirable performances. However, the  matrix completion  marginally outperforms  the hybrid design with an average performance gain of $0.2$ dB.

\begin{figure}[!t]
	\centering
	\includegraphics[width=9.55cm, height=6.90cm]{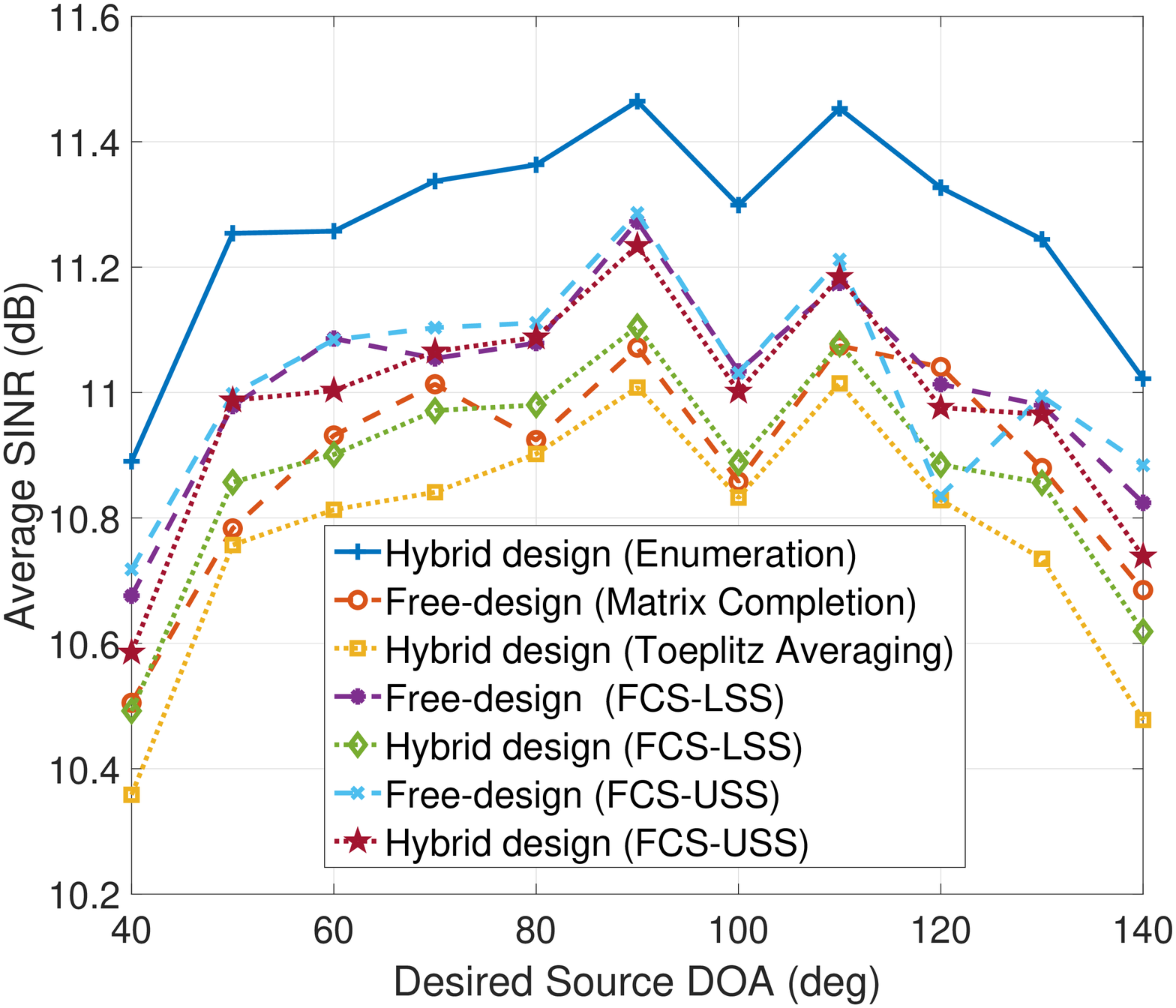}
	\caption{Average SINR performance of various sparse topologies against desired source DOA  for  $T=1000$ snapshots.}
	\label{robust_bp3}
\end{figure}

\begin{figure}[!t]
	\centering
	\includegraphics[width=9.55cm, height=5.50cm]{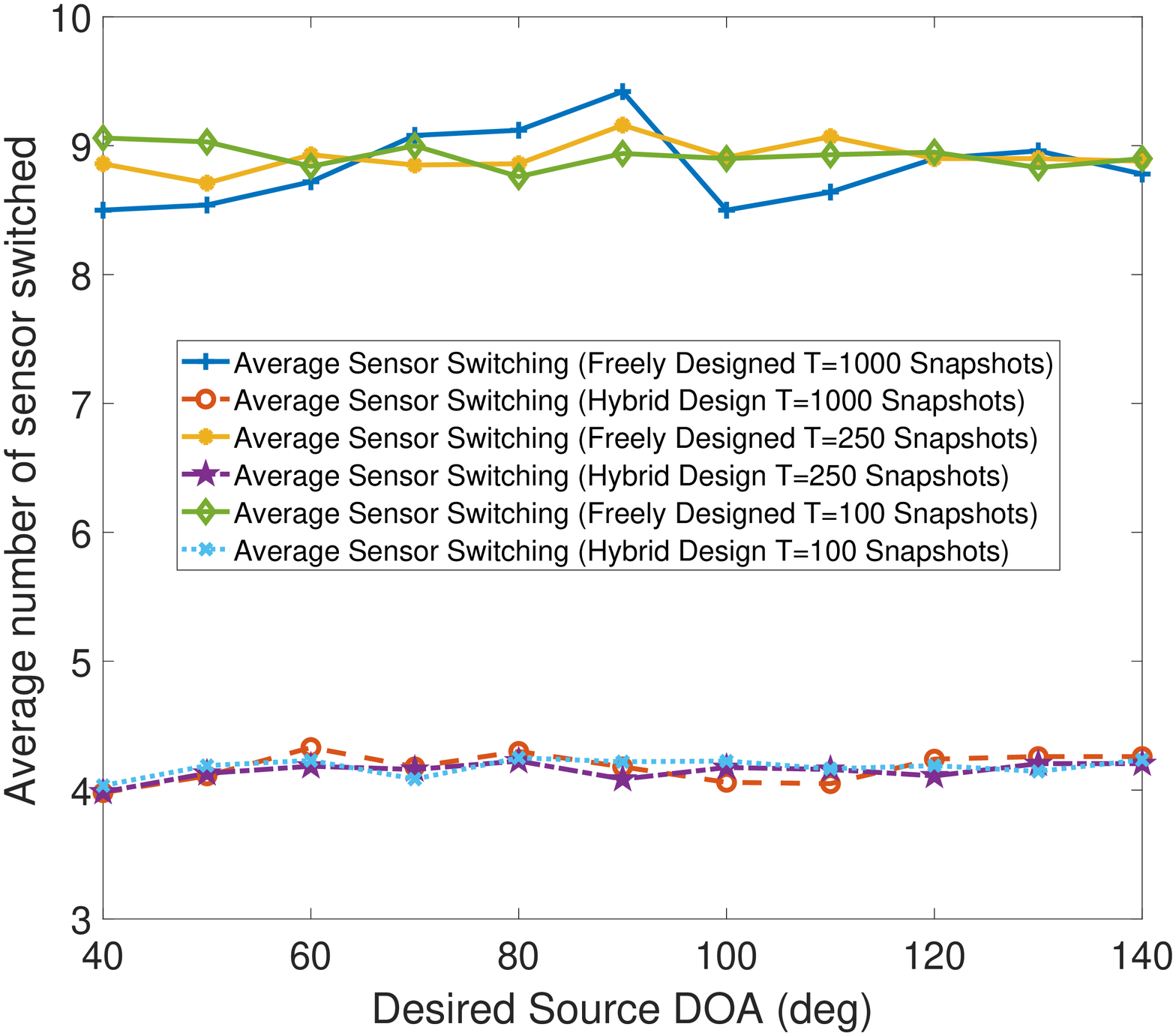}
	\caption{Sensor switching  comparison vs the  free-design and the hybrid design.}
	\label{robust_bp4}
\end{figure}

The performance curves  are re-evaluated by increasing the snapshots to $T=250$ and $T=1000$, as shown in Figs. \ref{robust_bp2} and \ref{robust_bp3}. With such increase, the performances of the proposed SCA using Toeplitz completion move closer to the performances of the FCM-USS benchmark. It is also noted that in contrast to lower snapshots ($T=100$), the  FCM-LSS benchmark  for higher samples ($T=1000$) offers superior average performance over SCA designs incorporating Toeplitz completion.   It is of interest to track the average antenna switching involved per trial for both the free-design and the hybrid design. Fig. \ref{robust_bp4} shows that freely designed array involves $9$ antenna switching per trial which is more than twice that of  the hybrid design ($4$ antenna switching per trial). It is also  noted that for the hybrid design, the maximum antenna switching is constrained to  $5$ antennas as the rest of $11$ sensors are prefixed. In this regard, the hybrid design    has  more efficient switching as it utilizes 80 percent (4/5) of the DOF as compared to the mere 55 percent (9/16) switching efficiency of freely designed arrays. 

\section{Conclusion}
Sparse array design for maximizing the beamformer output SINR is considered for a single source in an interference active environment. The paper addressed the problem  that the optimization of the array configuration requires full data correlation matrix which is not readily available in practice. Two different design approaches were considered; one assumes prefixed position of subset of sensors so as to provide full array augmentation, referred to as the hybrid-design approach,  whereas the other, which is referred to as free-design approach,   has no such restriction, and freely allocates all degrees of freedom to maximize the objective function.    It was  shown that the Toeplitz estimation of the autocorrelation at the missing spatial lags has a desirable performance. The SCA was proposed for both the freely designed  and hybrid designed arrays  to achieve MaxSINR  in polynomial run times with a reasonable trade off in  SINR. It was  shown that, in contrast to hybrid design, the matrix completion scheme does not require to pre-allocate sensor resources and, therefore,  offers more design flexibility and better SINR performance. This performance improvement is, however,  at the cost of increased computational complexity and finer parameter tuning as required to accomplish Toepltiz matrix completion. The simulation examples provided showed   that the performance of the proposed SCA algorithm incorporating  Toeplitz completion is agreeable with the established benchmark designs.

%\section*{References}

\bibliography{references}

\begin{thebibliography}{10}
\bibitem{doi:10.1002/cpa.20132}
D.~L. Donoho, For most large underdetermined systems of linear equations the
  minimal $l_1$-norm solution is also the sparsest solution, Communications on
  Pure and Applied Mathematics 59~(6)  797--829.
\end{thebibliography}

\end{document}